\documentclass[12pt,a4paper,twocolumn]{narms}

\usepackage{graphicx}
\usepackage{subfigure} 
\usepackage{psfig}
\usepackage{timesmt}   
\usepackage{chikako}   
\usepackage{latexsym}

\makeatletter
\renewcommand{\section}{\@startsection%
{section}%
{1}%
{0mm}%
{- \baselineskip}%
{0.15\baselineskip}%
{\normalfont\normalsize}}%
\renewcommand{\subsection}{\@startsection
{subsection}%
{2}%
{0mm}%
{-\baselineskip}%
{0.15\baselineskip}%
{\normalfont\normalsize}}%
\makeatother

\begin{document}

\title{Pattern formation by injection of air in a non-Brownian suspension}
\author{\large {C. Chevalier, A. Lindner \& E. Cl\'ement}\\
{\em Physique et M\'ecanique des Milieux H\'et\'erog\`enes, Ecole Sup\'erieure de Physique et de Chimie Industrielles }\\ \em {10 rue Vauquelin, 75231 Paris Cedex 05, France}\\
}
\date{}

\abstract{ABSTRACT: We study the injection of air in a granular suspension. We use a linear Hele-Saw cell filled with a suspension which is displaced by air, leading to a Saffman-Taylor (fingering) instability. For the suspension, we use an iso-dense mixture where the fluid and the particles have the same density. The volume fraction of particles can thus be adjusted over a wide range. We discuss the question of an effective rheology inside the cell as well as the pattern formation as a function of the granular compacity. We finally report results on the finger width for stable fingers and the thresholds for their destabilization.}
\maketitle
\frenchspacing

\section{INTRODUCTION}

Viscous fingering has received much attention as an archetype of pattern-forming systems and was stud-ied intensively for Newtonian fluids (Saffman \& Taylor 1958, Bensimon et al. 1986, Homsy 1987) and non-Newtonian fluids \shortcite{Lind02}. It is an important model system suited for a better understanding of fluid injection in porous media with important practical applications such as crude oil recovery.

Here, we consider a case where the fluid to be displaced is an isodense non-Brownian suspension. The volume fraction of the grains can be easily controlled and thus, the rheology of the suspension can be varied over a wide range: going from an effective viscous fluid up to a jammed dense paste. This system can be seen as a fundamental tool to understand many complex injection problems such as those existing in weakly consolidated porous media. Its potential interest would lie in a better understanding of many practical processes such as the stabilization of saturated soils by air injection or the consolidation of oil extraction wells.

A difficult question is still to understand the interfacial dynamics between a suspension and another fluid. Viscous fingering is mainly governed by the local viscosity at the finger tip and is also very sensitive to the presence of particles. We thus expect that studying the pattern formation will lead to a better understanding of particles dynamic at the interface between suspension and air (or another injection fluid).

The paper is organized as follows. In Sec.2 we will recall the basic equations for the Saffman-Taylor instability. Sec. 3 describes the set-up and experimental methods. In Sec. 4, the experimental results concerning the rheology of suspensions and the Darcy's law are presented and discussed. In Sec. 5, we interest ourselves at the finger shape as well as the finger width. Sec. 6 gives a summary of the obtained results

\section{THE SAFFMAN-TAYLOR INSTABILITY}

\subsection{\em Presentation of the instability}

\begin{figure}[htb]
        \begin{center}
            \includegraphics[height=2.8cm]{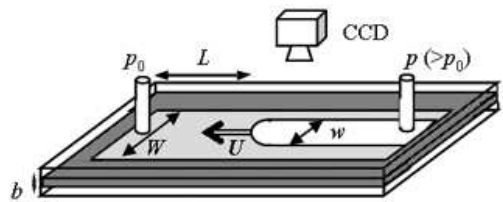}
            \caption{\small Schematic drawing of the experimental set-up.}
            \label{HS}
        \end{center}
    \end{figure}

The Saffman-Taylor instability is typically studied in a thin linear channel or Hele-Shaw cell (Fig. \ref{HS}). The width of the cell $W$ is chosen to be large compared to the channel thickness $b$. The cell is filled with a viscous fluid which is then pushed by air. The properties of the viscous fluid are its viscosity $\mu$, its surface tension $\gamma$ and its density $\rho$. The viscosity and the density of air are neglected.

When air pushes the viscous fluid (for example due to an imposed pressure gradient $\nabla P$), the interface between the two fluids destabilizes. This destabilization leads to the formation of a viscous finger of width $w$ propagating at velocity $U$. In general, one considers the relative finger width: $\lambda=w/W$.

\subsection{\em Averaged Darcy's law}

For Newtonian fluids, the motion in the Hele-Shaw cell is described by the two-dimensional gap averaged velocity field ${\bf u}$. It is given by the local Darcy's law, which relates the local pressure gradient to the fluid velocity.

Away from the interface the flow can be considered as uniform and one can relate the mean flow velocity $V$ to the pressure gradient $\nabla P$:
\begin{equation}
V=-\frac{b^{2}}{12\eta } \nabla P.
\end{equation}

\subsection{\em Finger selection}

The relative finger width $\lambda$ is determined by the capillary number $Ca = \eta U / \gamma$ which represents the ratio of viscous forces and capillary forces. The viscous forces tend to narrow the finger, whereas the capillary forces tend to widen it. One thus anticipates that the relative width of the viscous fingers decreases with increasing finger velocity. For large values of the capillary parameter, $\lambda$ reaches a limiting value of about half the channel width.

The control parameter of the instability is $ 1/B = 12 (W/b)^2 Ca$ with $W/b$ the aspect ratio of the Hele-Shaw cell. When scaled on $1/B$, measurements of $\lambda$ for different systems all fall on the same universal curve (Saffman \& Taylor 1958, McLean \& Saffman 1981).

\section{EXPERIMENTAL SET-UP}

We use polystyrene spheres of 40~$\mu$\/m in diameter in a modified silicone oil DC704 of viscosity $\eta_0$~=~59.5~mPa.s and surface tension $\gamma$~=~31.0~mN/m. The density $\rho$ of the fluid matches the one of the grains (1.05~g/cm$^3$) closely leading to what we call an iso-dense suspension. This allows us to control the fraction of grains over a wide range and no sedimentation occurs on the timescale of our experiments. Here we work with grain fractions (i.e. the amount of grains on the total amount of suspension) $\Phi$ up to 30~\%.

We worked in a linear Hele-Shaw cell consisting of two glass plates separated by a thin Mylar spacer (Fig.~\ref{HS}). These plates were horizontal and clamped together in order to obtain a regular thickness of the channel $b$ of 0.75~mm for a width $W$ of 4~cm. The cell was filled with the suspension and compressed air was used as the less viscous driving fluid.

Fingers or more complex patterns were captured by a CCD camera, coupled to a data acquisition card (National Instruments) and a computer. In particular, this allowed for measurements of the relative width $\lambda = w/W$ as a function of the velocity $U$ for stable fingers. For each grain fraction several experimental runs (between 20 and 30) were performed using different applied pressure drops leading to many different finger velocities.

The rheology of our suspensions was studied in a double Couette geometry (2x~0.5~mm) on a ThermoHaake rheometer.

\section{RESULTS: RHEOLOGY AND DARCY'S LAW}

\subsection{\em Rheology}

\begin{figure}[htb]
        \begin{center}
            \includegraphics[height=5cm]{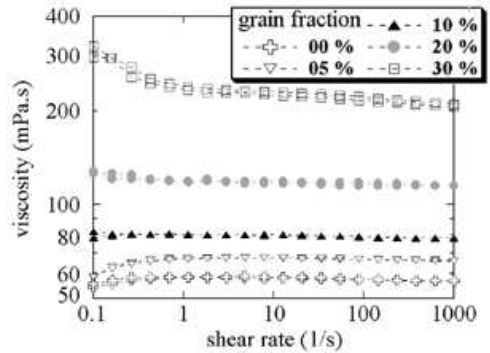}
            \caption{\small Rheological measurements: viscosity as a function of the shear rate.}
            \label{Rheo}
        \end{center}
    \end{figure}

Figure \ref{Rheo} shows the viscosity of the suspensions as a function of the shear rate $\dot{\gamma}$ for different grain fractions $\Phi$.

For all suspensions one observes Newtonian behavior over a wide range of shear rates (1~s$^{-1}$ $<$ $\dot{\gamma}$ $<$ 100 s$^{-1}$). Our fingering experiments are indeed performed in this range of shear rate and we can thus define a constant viscosity $\eta_{Rh}$ from the rheological measurements for each suspension. 

For large shear rates ($\dot{\gamma}$~$>$~100~s$^{-1}$) and high grain fraction ($\Phi$~=~30~\%) weak shear thinning is observed.

For low shear rates ($\dot{\gamma}$~$<$~1~s$^{-1}$) and high grain fractions a decrease of the viscosity is observed which could be due to the existence of a yield stress, whereas for low grain fractions an increase of viscosity is observed.

Note that these results are in qualitative and quantitative agreement with previous work \shortcite{Zar00}.

\subsection{\em Darcy's law}

Assuming the flow far away from the finger to be uniform, we expect that the classical Darcy's law (Equation 1) linking the gap averaged fluid velocity $V$ to the imposed pressure gradient $\nabla P$ in our Hele-Shaw cell remains valid.

The imposed pressure gradient is calculated by: $\nabla P = \Delta p / L$ where $\Delta p$ is the applied pressure drop and $L$ the distance between the finger tip and the exit of the cell. Mass conservation allows to obtain the velocity $V$ of the fluid far away from the interface from the finger velocity $U$ simply by using $V = \lambda U$ \shortcite{Lind00}.

\begin{figure}[htb]
       \begin{center}
            \includegraphics[height=5.2cm]{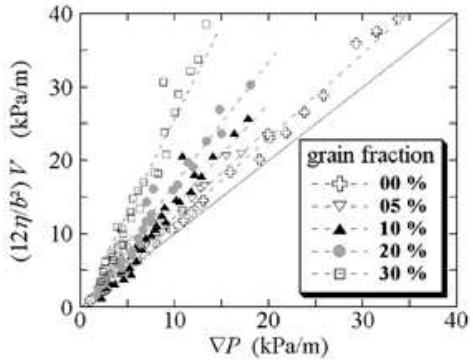}
            \caption{\small Darcy's law for the suspensions. The straight line represents Darcy's law expected in a Hele-Shaw cell.}
            \label{Dar}
        \end{center}
    \end{figure}

Following equation 1, results are described on Figure \ref{Dar} representing $V.(12 \eta_{Rh}/b^2)$ as a function of $\nabla P$. If one sompares with what is obtained from Darcy's law (Fig.~\ref{Dar} one observes systematic deviations towards higher slope values, which increase with increasing volume fraction. This might indicate that there exists an effective viscosity $\eta_C$ in the cell which is lower than $\eta_{Rh}$ obtained with the rheometer.

\subsection{\em Flow in the cell}

\begin{figure}[htb]
        \begin{center}
            \includegraphics[height=3cm]{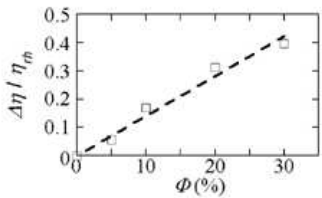}
            \caption{\small Relative viscosity difference as a function of grain concentration $\Phi$.}
            \label{Phi1}
        \end{center}
    \end{figure}

Figure \ref{Phi1} reports the relative difference of the viscosities:
 \begin{equation}
\frac{\Delta \eta}{\eta_{Rh}}(\Phi)=\frac{\eta_{Rh}(\Phi)-\eta_C (\Phi)}{\eta_{Rh}(\Phi)}
\end{equation}
as a function of $\Phi$. The cell viscosity $\eta_C$ is always lower than $\eta_{Rh}$ and the relative difference increases with the grain fraction.

These results could be qualitatively explained by a granular structuring of the flow in the thickness of the cell \shortcite{Lyon98}: on one side, the flow is no longer parabolic but evolves towards a plug flow with increasing grain fraction and, on the other side, the grain concentration is no longer homogeneous and becomes higher in the central region at low shear. It is clear that such a restructuring of the flow could lead to a lower average viscosity $\eta_C$.  Unfortunately, no direct link with an effective cell viscosity has been established so far.
 
\section{RESULTS: FINGERS AND PATTERNS}

\subsection{\em Experimental observations}

\begin{figure}[htb]
        \begin{center}
            \includegraphics[height=4cm]{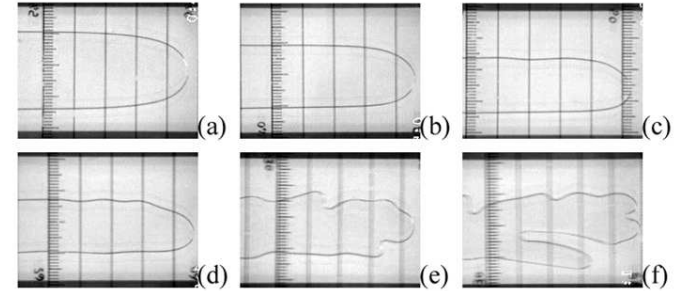}
            \caption{\small Typical evolution of the finger shape with increasing velocities: 0.3 (a), 2 (b), 5 (c), 10 (d), 12 (e) and 22 cm/s (f).}
            \label{Pat}
        \end{center}
    \end{figure}

Figures \ref{Pat}a,b,c,d,e,f present the typical evolution of patterns observed for increasing velocity, valid for most grain fractions ($\Phi$ up to 40~\%, not presented here).

For low velocities, the finger is stable and symmetrical with respect to the cell central axis (Fig. \ref{Pat}a). For higher velocity, it become asymmetric (Fig. \ref{Pat}b), although it remains stable (i.e. we could measure its width).

The first destabilization occurs on one side of the finger like a sinusoidal wave (Fig. \ref{Pat}c). Thereafter, for increasing velocity, this type of destabilization occurs on both sides (Fig. \ref{Pat}d).

At ever larger velocity, we observe classical destabilization of Saffman-Taylor fingers like side-branching (Fig. \ref{Pat}e) or tip-splitting (Fig. \ref{Pat}f).

\subsection{\em Finger width}

For stable fingers (Figs \ref{Pat}a, b), we measured the relative finger width $\lambda$. When representing $\lambda$ as a function of the control parameter $1/B$ of the instability one has to decide which viscosity to use.

\begin{figure}[htb]
        \begin{center}
            \includegraphics[height=4.8cm]{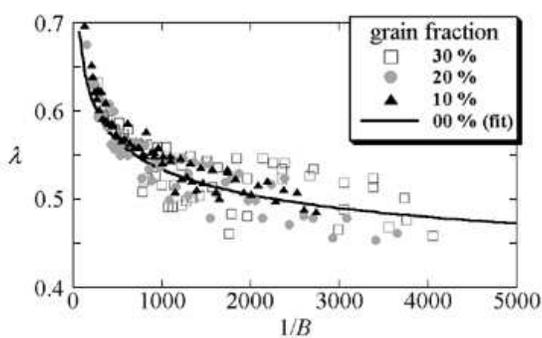}
            \caption{\small Finger width $\lambda$ as a function of the control parameter $1/B$ using $\eta_{Rh}$. The straight line represents a fit to the data for the pure fluid.}
            \label{ST1}
        \end{center}
    \end{figure}

\begin{figure}[htb]
        \begin{center}
            \includegraphics[height=4.8cm]{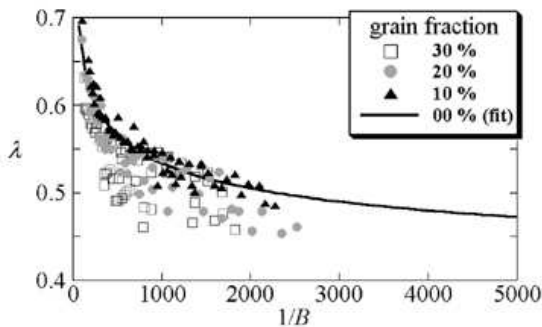}
            \caption{\small Finger width $\lambda$ as a function of the control parameter $1/B$ using $\eta_C$. The straight line represents a fit to the data for the pure fluid.}
            \label{ST2}
        \end{center}
    \end{figure}

Figures \ref{ST1} and \ref{ST2} represent $\lambda$ as a function of $1/B$ calculated with $\eta_{Rh}$ (Fig. \ref{ST1}) and with $\eta_{C}$ (Fig. \ref{ST2}). The line represents a fit to the data for the pure fluid and gives thus the master curve of the instability.

We observe an increasing dispersion of the results with grain fraction. However, on Figure \ref{ST1}, the data is scattered around the master curve whereas on Figure \ref{ST2}, all data seem to be below this curve.

As the finger width depends on the viscosity at the finger tip, one expects that a non-uniform flow and grain distribution will affect the finger width. Thus, one could interpret these results in terms of an effective viscosity at the finger tip which would vary around $\eta_{Rh}$ with $\eta_C$ being the lower limit.

\subsection{\em Instability thresholds}

\begin{figure}[htb]
        \begin{center}
            \includegraphics[height=3.5cm]{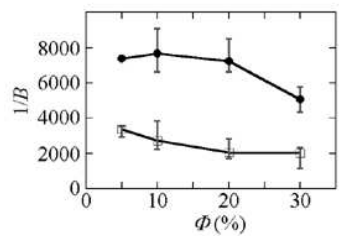}
            \caption{\small Instability thresholds: $\Box$, apparition of oscillations on one side of the finger; $\bullet$, apparition of side branching.}
            \label{Phi2}
        \end{center}
    \end{figure}

Following the evolution of the finger shapes (Figs \ref{Pat}), one observes several instability thresholds. On Figure \ref{Phi2}, the first threshold (apparition of oscillations on one side – Fig. \ref{Pat}c) and the threshold for side branching (Fig. \ref{Pat}e) are plotted. Here $\eta_C$ is used to obtain the value of $1/B_c$.

We observe little variations of the first threshold with $\Phi$ which is found to be $1/B_c$~$\approx$~2500 whereas in the case of pure fluid (i.e. $\Phi$~=~0), we observe stable fingers up to $1/B$~$\approx$~10,000.

Following the work of Bensimon et al. (1986), one can obtain a critical amplitude of the perturbations leading to a finger destabilization at a value $1/B_c$~=~2500 for a finite amplitude $A_c$~$\approx$~5$\sim$20~$\mu$m. 
Furthermore, this work predicts the destabilization to be independent on the wavelength of the perturbation. These predictions are in good agreement with our experimental data if we consider the size of the grains (40 $\mu$m) to be of the order of $A_c$ and the threshold to be weakly dependent of fraction of the grains (i.e. the concentration of grains at the finger tip) which should be proportional to the wavelength of the perturbation. To test this further experiments are planned changing the particle diameter.

\section{CONCLUSION AND PERSPECTIVES}

In summary we have reported results on the injection of air in granular suspensions. The dynamics of the air/suspension interface can be described by a Saffman-Taylor like instability.
Rheological measurements and the use of Darcy's law for flow in the Hele-Shaw cell lead to different values for the viscosities of the suspensions. This could be explained by a non-homogeneous flow in the cell.

Furthermore we report the evolution of the observed patterns with finger velocity. For stable fingers, an increasing scatter of the data of the relative finger width with the grain fraction was observed, which might be linked to the local viscosity at the finger tip.

Finally, an analysis of the instability thresholds leads us to underline the importance of the grain size for the destabilization of the fingers.

In further studies of the dynamics of the interface between a pure fluid and a suspension, two cases seem to be of interest: the case of dense suspensions ($>$~50~\%) and the case of miscible flow where the pushing pure fluid and the suspending one are identical.

We thank M. Clo\^itre (ESPCI, Paris) for helping us with the rheological measurements and Daniel Bonn (ENS, Paris) for helping us with the experimental set-up.


\bibliographystyle{chikako}      

\newpage
\begin{thebibliography}{}

\bibitem[\protect\citeauthoryear{Bensimon}{Bensimon et al.}{1986}]{Ben86}
Bensimon, D. et al. 1986.
\newblock Viscous flows in 2 dimensions.
\newblock {\em Reviews of Modern Physics\/}~{ 58}: 977--999.

\bibitem[\protect\citeauthoryear{Homsy}{Homsy}{1987}]{Hom87}
Homsy, G.M. 1987.
\newblock Viscous fingering in porous media.
\newblock {\em Annual Review of Fluid Mechanics\/}~{ 19}: 271--311.

\bibitem[\protect\citeauthoryear{Lindner}{Lindner et al.}{2000}]{Lind00}
Lindner, A. et al. 2000.
\newblock Viscous fingering in a shear-thinning fluid.
\newblock {\em Physics of Fluids\/}~{ 12}: 256--261.

\bibitem[\protect\citeauthoryear{Lindner}{Lindner et al.}{2002}]{Lind02}
Lindner, A. et al. 2002.
\newblock Viscous fingering in non-newtonian fluids.
\newblock {\em Journal of Fluid Mechanics\/}~{ 469}: 237--256.

\bibitem[\protect\citeauthoryear{Lyon}{Lyon \& Leal}{1998}]{Lyon98}
Lyon, M.K. \& Leal, L.G. 1998.
\newblock An experimental study of the motion of concentrated suspensions in
  two-dimensional channel flow. part 1. monodisperse systems.
\newblock {\em Journal of Fluid Mechanics\/}~{ 363}: 25--56.

\bibitem[\protect\citeauthoryear{McLean}{McLean \& Saffman}{1981}]{Lean81}
McLean, J.W. \& Saffman, P.G. 1981.
\newblock The effect of surface ten-sion on the shape of fingers in a hele-shaw
  cell.
\newblock {\em Journal of Fluid Mechanics\/}~{ 102}: 455--469.

\bibitem[\protect\citeauthoryear{Saffman}{Saffman \& Taylor}{1958}]{ST58}
Saffman, P.G. \& Taylor, G.I. 1958.
\newblock The penetration of a fluid into a porous medium or hele-shaw cell
  containing a more viscous liquid.
\newblock {\em Proceedings of the Royal Society of London A\/}~{ 245}:
  312--329.

\bibitem[\protect\citeauthoryear{Zarraga}{Zarraga et al.}{2000}]{Zar00}
Zarraga, I.E. et al. 2000.
\newblock The characterization of the total stress of concentrated suspensions
  of noncolloidal spheres in newtonian fluids.
\newblock {\em Journal of Rheology\/}~{ 44}: 185--220.

\end{thebibliography}

\end{document}